\documentclass[%
reprint,
superscriptaddress,
 amsmath,amssymb,
 aps,
]{revtex4-2}

\usepackage{graphicx}% Include figure files
\usepackage{dcolumn}% Align table columns on decimal point
\usepackage{bm}% bold math
\usepackage{hyperref}% add hypertext capabilities
\usepackage{todonotes}
\usepackage{siunitx}
\usepackage[normalem]{ulem}
\newcommand{\rv}[1]{\textcolor{black}{#1}}
\newcommand{\rvv}[1]{\textcolor{black}{#1}}
\begin{document}

\preprint{APS/123-QED}

\title{\rv{Emergence of capillary waves in miscible \rvv{co-flowing} fluids}}% Force line breaks with \\

\author{Alessandro Carbonaro}
\affiliation{Laboratoire Charles Coulomb (L2C), UMR 5221 CNRS-Universitè de Montpellier, F-34095 Montpellier, France}%
\author{Giovanni Savorana}%
\altaffiliation[Currently at ]{Institute of Environmental Engineering, ETH Zurich, 8093, Zurich, Switzerland}%
\affiliation{Laboratoire Charles Coulomb (L2C), UMR 5221 CNRS-Universitè de Montpellier, F-34095 Montpellier, France}%
\author{Luca Cipelletti}
% \homepage{http://www.Second.institution.edu/~Charlie.Author.}
\affiliation{Laboratoire Charles Coulomb (L2C), UMR 5221 CNRS-Universitè de Montpellier, F-34095 Montpellier, France}%
\affiliation{Institut Universitaire de France, F-75231 Paris, France}
\author{Rama Govindarajan}
\affiliation{International Centre for Theoretical Sciences, Tata Institute of Fundamental Research, Shivakote,
Bengaluru 560089, India}
\author{Domenico Truzzolillo}
% \homepage{http://www.Second.institution.edu/~Charlie.Author.}
\email{domenico.truzzolillo@umontpellier.fr}
\affiliation{Laboratoire Charles Coulomb (L2C), UMR 5221 CNRS-Universitè de Montpellier, F-34095 Montpellier, France}%


\date{\today}% It is always \today, today,

\begin{abstract}
We \rv{show that} capillary waves \rv{can exist at the} the boundary between miscible co-flowing fluids. We unveil that the interplay between transient interfacial stresses and confinement drives the progressive transition from the well-known inertial regime, characterized by a \rv{frequency independent} wavenumber, $k\sim\omega^{0}$, to a capillary wave scaling, $k\sim\omega^{2/3}$, \rv{unexpected for miscible fluids}. This allows us to measure the effective interfacial tension between miscible fluids and its rapid decay on time scales never probed so far, which we rationalize with a model going beyond square-gradient theories. \rv{Our work potentially opens a new avenue to measure transient interfacial tensions at the millisecond scale in a controlled manner.}
\end{abstract}

\maketitle
%\tableofcontents
%%INTRO%%

Capillary waves are small disturbances propagating at the interface between two fluids, \rv{with interfacial tension acting as the only relevant restoring force}~\cite{lambHydrodynamics1932,rayleighProgressiveWaves1877,chouCapillaryWaveScattering1995}. 
They are very common in Nature \cite{lambHydrodynamics1932} but can also be easily produced artificially \cite{bobbCapillaryWaveMeasurements1979,behrooziDirectMeasurementAttenuation2001}. 
Notably, capillary waves have proved to be a valuable tool to measure interfacial or surface tension between immiscible fluids \cite{sohlNovelTechniqueDynamic1978}, since their dispersion relation links the tension $\Gamma$ to the wavenumber $k$ and frequency $\omega$ of the waves.
Recently, Hu and Cubaud \cite{huViscousWaveBreaking2018} have shown that propagating modes can be excited in viscosity-stratified flows confined in microfluidic channels, and that some of the waves observed for coflowing immiscible liquids follow the capillary wave dispersion predicted for inviscid fluids, $k\sim \omega^{2/3}$ \cite{lambHydrodynamics1932}. 
Hence, even in confined flows, capillary waves can emerge when interfacial tension dominates over other stabilizing mechanisms such as viscous dissipation.
In the flow of \rv{\emph{miscible} fluids, by contrast, interfacial stresses are typically negligible compared to viscous dissipation, such that their very  existence is debated by an increasing number of research groups~\cite{bierNonequilibriumInterfacialTension2015,truzzolilloNonequilibriumInterfacialTension2016c,suzukiExperimentalStudyMiscible2020,gowdaFormationColloidalThreads2021,vorobevNonequilibriumCapillaryPressure2021,lacazeTransientSurfaceTension2010}}. An ``inertial'' regime characterizes the emerging instability, with $k\sim \omega^0$ \cite{huViscousWaveBreaking2018}: the variation of $\omega$ is solely dictated by a change in phase velocity, while the wavelength stays practically unaltered when the flow rate of either fluid is varied. This regime has also been observed for high Reynolds numbers in immiscible fluids, where it delimits the short wavelength limit \cite{huViscousWaveBreaking2018} of the stability diagram, suggesting a maximum wavenumber dictated only by the rate of energy dissipation of the waves. %On that account, the inertial regime in confined flows displays smaller wavelengths compared to capillary waves, that, by contrast, are considered as the smallest interfacial modes excited in open waters \cite{leblondWavesOcean1978}. 
Hence, dissipation, confinement, and interfacial tension are key to \rv{understanding flow instabilities} and the transition from the inertial to the capillary regime.\\ 
\\ 
\rvv{To date, the presence of capillary waves in miscible fluids has only been inferred via light scattering in static conditions and for nearly critical fluids \cite{cicutaCapillarytobulkCrossoverNonequilibrium2001}, whose miscibility can be easily tuned by changing temperature, which however represent a limited subset of fluid pairs to study.}
\rvv{The direct observation of} capillary waves in fluids, \rvv{miscible in all proportions at any temperature below their boiling point,} has not been reported so far, most likely due to the rapidly decaying tension at their boundaries \cite{kortewegFormeQuePrennent1901,pojmanEvidenceExistenceEffective2006b,cicutaCapillarytobulkCrossoverNonequilibrium2001,smithTransientInterfacialTension1981,truzzolilloEquilibriumSurfaceTension2014,truzzolilloNonequilibriumInterfacialTension2016c}. \rv{Microfluidic devices provide an ideal playground to test whether interfacial stresses can transiently impact wave propagation, since the behavior of the interface between two fluids can be explored on time scales as short as a few tens of milliseconds since the fluids are brought in contact.}  %formation at the interface between coflowing liquids in microfluidic devices can be a very fast process, with just a few tens of milliseconds elapsing from the first contact between fluids to the instability occurrence, microfluidics provides an ideal playground to test whether interfacial stresses due to large concentration gradients can transiently impact wave propagation. 
In this letter, we investigate the propagation of shear flow instabilities appearing when \rv{two miscible fluids}, water and glycerol, flow side-by-side in rectangular microchannels. We unveil the existence of capillary waves, \rv{with} two distinct branches of wave dispersion, \rv{and we measure the effective interfacial tension} at the liquid-liquid boundary over time scales never probed so far, \rv{opening a new avenue to measure transient interfacial tensions in a controlled manner at very short times}.

\begin{figure}[htbp]
   %\Requires \usepackage{graphicx}
 \includegraphics[width=\linewidth]{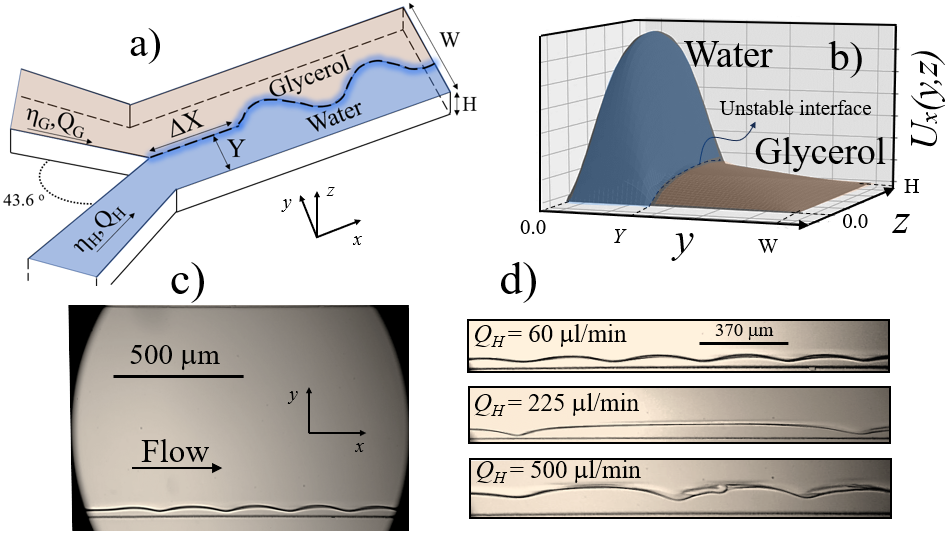}
 \caption{
 Experimental setup and flow configuration. \textbf{a}: Scheme of the Y-junction chip to generate co-flowing fluids. \textbf{b}: Characteristic 3D-base flow \rv{in the main duct,} for water and glycerol. \textbf{c}: Entire field of view for ($Q_{G}$,$Q_{H}$)=(18 $\mu$l/min, 60 $\mu$l/min) and $W=1$mm. \textbf{d}:  Interfacial instabilities for three different \rv{$Q_H$, with fixed} $Q_{G}$ = 18 $\mu$l/min. Note that $\lambda(Q_{H}$) is non-monotonic.
 }\label{fig1}
\end{figure}

We generated and visualized interfacial waves by pumping deionized water and glycerol in polydimethylsiloxane (PDMS) microchannels, as sketched in Fig. \ref{fig1} (a) (\cite{SeeSupplementalMaterial}). Most of the experiments have been carried out in a channel whose main duct has height $H=0.1$ mm and width $W=1$ mm. Further experiments to inspect the role of lateral confinement were performed in a narrower channel, $W=0.25$ mm, with the same height and Y-geometry. 
The two fluids flow parallel to each other and the interface is vertical, parallel to the gravity direction $\hat{z}$, thus ruling out any effect due to density mismatch \rv{in the observed region}.%, at least in the first portion of the channel, where the two fluids stay in parallel co-flow and do not re-orient themselves to minimize gravitational energy.

The instability is observed via optical microscopy at its onset, and it is characterized by recording the interface dynamics at fixed flow rate of glycerol $Q_{G}$ and varying the flow rate of water from $Q_{H}=0$ $\mu$l/min to several hundreds of $\mu$l/min, depending on $Q_{G}$. We repeated this protocol in the range  $3 \mu l/min\leq Q_G\leq 25 \mu l/min$. The instability appears above a critical flow rate of water $Q^c_{H2O}$ that depends marginally on $Q_G$ (\cite{SeeSupplementalMaterial}). The typical base flow is sketched in Fig. \ref{fig1}-b, with water being the faster moving fluid. The velocity profiles in absence of diffusion change curvature at the interface, located at $y=Y$. This configuration allows to link the stream-wise position $\Delta X$ of the instability onset and the diffusion time $t_c$, since $t_c=\Delta X/U(Y,H/2)$, where $U(Y,H/2)$ is the base flow velocity of the interface at the imaged plane $z=H/2$. 

As shown in Figs \ref{fig1} (c,d), the interface can be visualized thanks to the \rv{difference in refractive index between water and glycerol}. Here, both the fluids flow from left to right, with glycerol occupying the upper part of the channel. The shape of the wave front in each frame is reconstructed by tracking the position $Y$ of the interface in the direction orthogonal to the flow direction, $\hat{x}$, thus obtaining a curve in the $x-y$ plane that closely represents the contour of the deformed interface~\cite{SeeSupplementalMaterial}. The spatio-temporal averages and the respective standard deviations of the wavelength, the phase velocity in the laboratory frame and the amplitude are directly extracted from image analysis~\cite{SeeSupplementalMaterial}. 

Figure \ref{fig1}-(c) and (d) set the scene for describing the general phenomenology of the observed co-flow instabilities. In (c) we show the entire field of view for one single set of flow rates. The instability develops close to the wall at $y=0$. \rv{Indeed, in all experiments} the unperturbed interface stays in the range 30 $\mu$m $\leq Y\leq$ 320 $\mu$m. In (d) we show three frames depicting the emerging instability at three selected flow rates of water for the same $Q_G$=18 $\mu$l/min. Remarkably, the wavelength of the instability does not vary monotonically upon increasing $Q_{H}$: the average wavelength $\lambda$ of the instability first increases with $Q_{H}$, reaching very large values (up to $\lambda>W$) for intermediate $Q_{H}$, and finally declines when the water rate is increased further. In none of the cases we observe a spatio-temporal evolution of the instability amplitude in the accessible field of view ($1.6$ mm downstream of the onset): almost instantaneously the observed waves reach a steady amplitude, pointing to a non-linear saturation in the instability production with mode-coupling at play \citep{landauCourseTheoreticalPhysics1987}. This has been already underlined in \cite{huViscousWaveBreaking2018} for other pairs of fluids. 

\begin{figure}[htbp]
   %\Requires \usepackage{graphicx}
 \includegraphics[width=\linewidth]{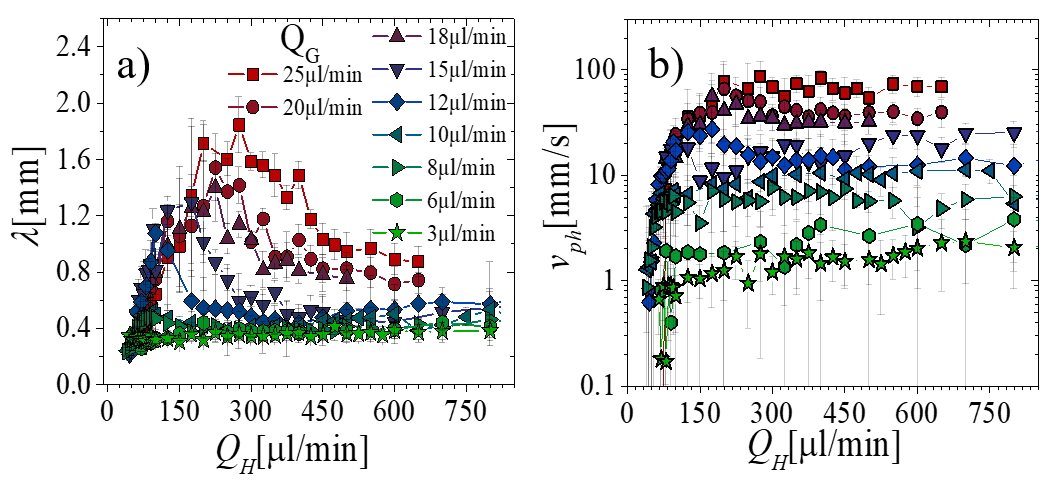}
 \caption{Wavelength $\lambda$ (\textbf{a}) and phase velocity $v_{ph}$ (\textbf{b}) of the interfacial waves \rv{\textit{vs}} the flow rate of water $Q_{H}$, for different flow rates of glycerol as indicated in (\textbf{a}). \rv{Same symbols} in both panels. Error bars are spatio-temporal standard deviations~\cite{SeeSupplementalMaterial}.}\label{fig2}
\end{figure}

Figures \ref{fig2}-a,b show the wavelength $\lambda$ and the phase velocity of the waves in the frame of the unperturbed interface, $v_{ph}=v_{ph}^{lab}-U(Y,H/2)$, with $U(Y,H/2)$ computed analytically or retrieved from the interface position \cite{SeeSupplementalMaterial}. Subtraction of the base flow velocity at $y=Y$ eliminates the Doppler shift caused by the underlying fluid motion \cite{dennerNumericalTimestepRestrictions2015,giamagasPropagationCapillaryWaves2023}\rv{; however,} our results do not change qualitatively if \rv{no subtraction is performed}, as in \cite{huViscousWaveBreaking2018}. 
As suggested by Fig \ref{fig1}-d, \rv{$\lambda$} develops a maximum, which grows in magnitude and shifts to higher rates of water for increasing rate of glycerol, Fig. \ref{fig2}-a. By contrast, the phase velocity initially increases sharply right above the critical rate of water $Q^c_{H}$ for the instability (Fig. \ref{fig2}-(b) and \cite{SeeSupplementalMaterial}), and then \rv{nearly plateaus, a behavior seen for all the investigated $Q_{G}$}. 
%Since the phase velocity of the waves does not show a behavior affine to that of the wavelength, we can already prefigure the existence of branches of the dispersion relation of this instability, with some values of the wavenumber $k=2\pi/\lambda$ characterized by two emerging frequencies. 
Similarly to $v_{ph}$, the amplitude of the instability exhibits two regimes with a rapid increase followed by a nearly constant plateau as a function of $Q_{H}$~\cite{SeeSupplementalMaterial}. Quite generally, the wavelength, the phase velocity and the amplitude~\cite{SeeSupplementalMaterial} \rv{increase systematically with $Q_{G}$ at all} $Q_{H}>Q^c_{H}$.

\begin{figure}[htbp]
   %\Requires \usepackage{graphicx}
 \includegraphics[width=\linewidth]{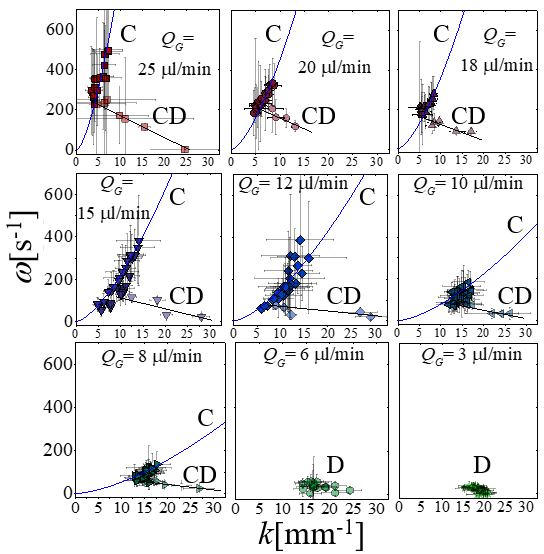}
 \caption{
 Dispersion relation for all $Q_{G}$ from the highest rate (25 $\mu$l/min) to the lowest one (3 $\mu$l/min). Blue lines are fits \rv{of Eq.~\ref{capillary} (capillary regime) to the \textit{C}-branch, which vanishes for $Q_{G}<8$ $\mu$l/min. Error bars are computed propagating the spatio-temporal variances measured for $\lambda$ and $v_{ph}^{lab}$.}  
 }\label{fig3}
\end{figure}

Figure \ref{fig3} shows the dispersion relation $\omega(k)=v_{ph}(k)k$ of the waves, with $k=2\pi/\lambda$. Two distinct branches \rv{are seen} for all rates of glycerol for which $\lambda(Q_H)$ is non-monotonic. \rv{The first branch, which we term \textit{C} (capillary), corresponds to the highest $Q_H$, with frequency increasing with $k$. In the second branch, termed \textit{CD} (capillary-dissipative), $\omega$ decreases with $k$ at high $Q_{G}$, becoming almost $k$-independent and eventually limited over a very small range of $k$ as $Q_{G}$ is decreased to 3 $\mu$l/min, where the \textit{C}-branch is absent and only \textit{D} (dissipative) waves are seen. See~\cite{SeeSupplementalMaterial} for a stability diagram in the $Q_G$-$Q_H$ plane, showing the \textit{C},\textit{CD} and \textit{D} regions. The rationale for the chosen terminology is discussed below.}

\begin{figure}[htbp]
   %\Requires \usepackage{graphicx}
 \includegraphics[width=\linewidth]{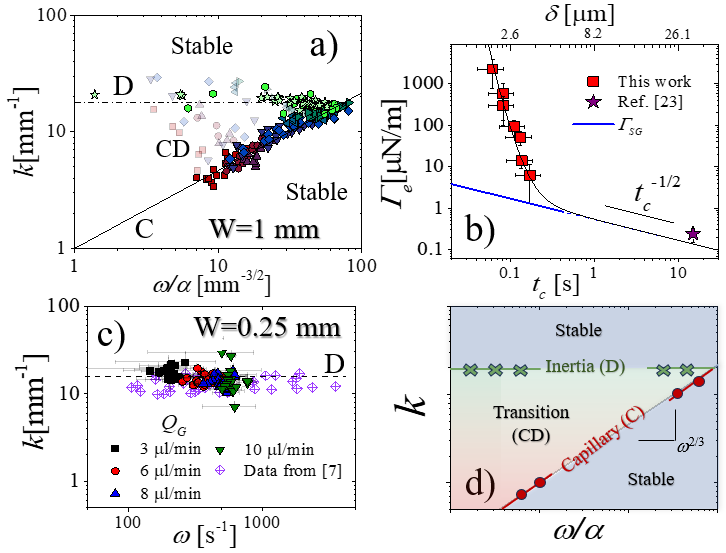}
 \caption{\textbf{a}: $k$ \textit{vs} scaled frequency $\omega/\alpha$ for the waves observed in the widest channel ($W=1$ mm). Same color code as in Fig. \ref{fig1}. The shadowed points are those belonging to the \textit{CD}-branches observed for those $Q_{G}$ where the \textit{C}-branch exists, and where the amplitude depends on the interface position~\cite{SeeSupplementalMaterial}. \rv{Error bars are omitted for clarity.} \rv{The solid line is the best power-law fit to the \textit{C}-branch detailed in the text}. The dashed horizontal line marks \rv{the wave number}, $k_D=18 mm^{-1}$, obtained by averaging the data for $Q_{G}=$6 $\mu$l and $Q_{G}=$3 $\mu$l. \textbf{b}: $\Gamma_e$ \textit{vs} $t_c$ (bottom axis) and interface thickness (top axis). \rv{Black solid line: best fit of Eq. \ref{efftens_tot}. Thick blue line: $\Gamma_{SG}$ computed as detailed in the text and in~\cite{SeeSupplementalMaterial}}
 \textbf{c}: Dispersion relation $k(\omega)$  for the waves observed in the narrowest channel ($W=0.25$ mm). \textbf{d}:
 Sketch of the general stability diagram.}\label{fig4}
\end{figure}

The \textit{C}-branch is strongly reminiscent of the capillary dispersion observed in immiscible co-flowing liquids \cite{huViscousWaveBreaking2018}: \rv{indeed, this branch is} well fitted by a capillary wave-type dispersion \cite{lambHydrodynamics1932} (blue lines in Fig. \ref{fig3}):
\begin{equation}\label{capillary}
\omega=\alpha k^{3/2}    
\end{equation}
with $\alpha$ that decreases for decreasing $Q_{G}$, until the \textit{C}-branch vanishes and only \textit{D}-waves remain. This scenario is quite surprising since capillary waves in confined shear flows were reported only for immiscible fluids, while miscible ones only showed an inertial regime with a wavenumber independent of frequency \cite{huViscousWaveBreaking2018}. 
To validate our fitting procedure we report in Fig. \ref{fig4}-a the wavenumber \textit{vs} the rescaled frequency $\omega/\alpha$ and we fit the entire set of points belonging to the \textit{C}-branches with a power-law $k=C_0(\omega/\alpha)^{\beta} $\rv{, obtaining $\beta=0.67\pm 0.05$, in excellent agreement with the capillary wave scaling $\beta_c=2/3$} \cite{lambHydrodynamics1932}.
From the best fit of Eq. \ref{capillary} to the \textit{C}-branches we computed the effective tension between water and glycerol for each $Q_{G}$, using the known relationship $\alpha=[\Gamma_e/(\rho_H+\rho_G)]^{1/2}$, \rv{with $\rho_H$, $\rho_G$} the densities of water and glycerol, respectively \cite{lambHydrodynamics1932,huViscousWaveBreaking2018}. 

Figure \ref{fig4}-b shows $\Gamma_e$ as a function of the average time $t_c=\Delta X/\langle v_{int}\rangle$ that the two fluids spend side-by-side until the instability takes place, where $\langle v_{int}\rangle$ is the unperturbed speed of the interface averaged over the rates $Q_{H}$ belonging to the \textit{C}-branch. \rv{This} averaging is sound, since \rv{$v_{int}$} depends mainly on $Q_{G}$ and only very marginally on $Q_{H}$~\cite{SeeSupplementalMaterial}. \rv{We also show $\Gamma_e$ obtained from the deformation of spinning droplets of water in a reservoir of glycerol~\cite{carbonaroUltralowEffectiveInterfacial2020a},} 15 s after the first contact between the fluids. The effective interfacial tension shows a very rapid decrease for increasing contact time,  from 2.3 mN/m at $t_c=$ 61 ms down to 6.3 $\mu$N/m at $t_c=$0.17 s, decaying way faster than the $\Gamma_e\sim t_c^{-1/2}$ scaling predicted for small gradients, where the square gradient (SG) expansion of the free energy is supposed to hold \cite{smithTransientInterfacialTension1981,petitjeansPetitjeanTensionSurface1996a}\rv{, eventually reaching much lower values than those reported previously\cite{petitjeansPetitjeanTensionSurface1996a}, see~\cite{SeeSupplementalMaterial}.} %The values that we obtain are in stark contrast with spinning drop tensiometry data reported previously for the same liquids: in \cite{petitjeansPetitjeanTensionSurface1996a} $\Gamma_e=0.58$ mN/m after about hundred of seconds from contact, a result leveraging on the assumption that water droplets spinning in a reservoir of glycerol evolve through a series of quasi-equilibrated states during their continuous elongation. The present results question this assumption and call for a different explanation.\\

We find that our entire set of data, including the data-point measured in \cite{carbonaroUltralowEffectiveInterfacial2020a}, is very well fitted (Fig. \ref{fig4}-b) by a model that some of us developed in \cite{truzzolilloNonequilibriumInterfacialTension2016c}, where higher order terms in the gradient expansion of the free energy are taken into account. The full expression for $\Gamma_e$ reads  
\begin{equation}\label{efftens_tot}
\Gamma_e(t_c)=\Gamma_{SG}(t_c)+\Gamma_0\left[\cosh\left(\frac{L}{\delta(t_c)}\right)-\frac{L^2}{2\delta^2(t_c)}-1\right],
\end{equation}
where \rv{$\Gamma_{SG}(t_c)=\int_{-\infty}^{+\infty}k_2(c)(\nabla c)^2dz$ is the square gradient contribution to the tension.
%\begin{equation}\label{efftens_GS}
%\Gamma_{SG}(t_c)=\int_{-\infty}^{+\infty}k_2(c)(\nabla c)^2dz
%\frac{RTa^2}{V_m\delta(t_c)}\left(\frac{2}{3}+\frac{\chi_{wg}}{6}\right)
%\end{equation}
Here $c\equiv c(z,t_c)$ is the local molar fraction of water, $z$ is the coordinate normal to the interface, and $k_2(c)=\frac{RTa^2}{V_m}\left\{\frac{\chi_{wg}}{2}+\frac{1}{\left[3c(1-c)\right]}\right\}$ is the square gradient coefficient computed on-lattice for a regular solution in presence of a locally flat interface \cite{truzzolilloNonequilibriumInterfacialTension2016c}, }% This reads}
%\begin{equation}\label{k2}
%k_2(c)=\frac{RTa^2}{V_m}\left\{\frac{\chi_{wg}}{2}+\frac{1}{\left[3c(1-c)\right]}\right\}
%\end{equation}}
where $R$ is the ideal gas constant, $a=0.45$ nm and $V_m=45.5$ ml/mol are the average \rv{diameter and molar volume} of the fluid molecules, respectively, and $\chi_{wg}=-0.96$ is the Flory interaction parameter for water and glycerol \cite{truzzolilloNonequilibriumInterfacialTension2016c}.
\rv{The second term on the r.h.s of Eq. \ref{efftens_tot} emerges uniquely from terms of orders higher than $O(\nabla c)^2$ in the expansion of the free energy density \cite{truzzolilloNonequilibriumInterfacialTension2016c} where} $\Gamma_0=\varepsilon\delta(t_c)$ sets the scale of the contribution to the effective tension due to large gradients at short times, via the energy density $\varepsilon$ and the interface thickness $\delta(t_c)$. $L$ is a characteristic length setting the \rv{value of $\delta$ below which terms of the free energy gradient expansion of order higher than $O(\nabla c)^4$ starts affecting the tension.} \rv{To obtain $\Gamma_{SG}$ and $\delta$, we computed $c(z,t_c)$ by solving the 1D diffusion equation, taking into account the concentration dependence of the mutual diffusion coefficient $D_{wg}(c)$ in water-glycerol mixtures~\cite{SeeSupplementalMaterial}. We obtained $\Gamma_{SG}=\frac{\gamma}{\sqrt{t_c}}$ with $\gamma=0.531$ $\mu$N/ms$^{1/2}$ at the timescales relevant for our microfluidics experiments \cite{SeeSupplementalMaterial} (blue line in  Fig. \ref{fig4}-b)). At short $t_c$, $\Gamma_{SG}$ largely underestimates the experimental $\Gamma_e$ and fails to capture its time dependence, while it extrapolates fairly well to the point measured in \cite{carbonaroUltralowEffectiveInterfacial2020a}.} 
%We further assume that the interface thickness $\delta(t_c)$ increases in time due only %to transverse diffusion, $\delta(t_c)=\sqrt{2D_{wg}t_c}$, with $D_{wg}=1.4 \times %10^{-11} m^2/s$ the diffusion coefficient of water in glycerol %\cite{carbonaroUltralowEffectiveInterfacial2020a,derricoDiffusionCoefficientsBinary2004a}.

\rv{To fit the entire set of data with Eq.~\ref{efftens_tot}, we defined $\delta(t_c) \equiv \delta_{\nabla}(t_c)$, the thickness of the region with steepest gradient, delimited by the two points where $\left|\partial^2 c/\partial z^2\right|$ is maximum, and  checked that alternative definitions of $\delta$ do not change the scenario described below~\cite{SeeSupplementalMaterial}. We obtained $\delta(t_c)=\Xi\sqrt{t_c}$ with $\Xi=8.24$ $\mu$m/s$^{1/2}$. (See \cite{SeeSupplementalMaterial} for the values of $L$ and $\varepsilon$ obtained from the fit and for a comparison with previously reported data). %We have further tested different proxies \cite{SeeSupplementalMaterial} for $\delta(t_c)$ and show that the quality of the fit is not affected by a different choice for $\delta(t_c)$, which only impacts $L$ and $\varepsilon$. 
The quantity $t_c^*=L^2$/$\Xi^2=$ 19.7 $\pm$ 5.2 s is independent of the definition chosen for $\delta$ and sets a characteristic time scale. For $t_c > t_c^*$, the leading correction to the square gradient expression scales as $ t_c^{-3/2}$. For $t_c<t_c^*$ higher order terms set in and $\Gamma_e(t_c)$ increases as $\sqrt{t_c}\cosh(\sqrt{t_c^*/t_c})$ for $t_c \rightarrow 0$.
%For the chosen proxy of $\delta(t_c)$ we obtain \rv{$L=36.6\pm 4.8$ $\mu$m and $\varepsilon=(54\pm 30)\times 10^{-6}$ Pa}.
} 
%Quite intriguingly, if we assume that the energy density $\varepsilon$ is dominated by excluded volume interactions and that local equilibrium \cite{kjelstrupNonequilibriumThermodynamicsHeterogeneous2008} may be invoked to use fluctuation-dissipation scaling as crude approximation to estimate the correlation length associated to this density, we obtain \rv{$l_{D}=(k_BT/\epsilon)^{1/3}=4.21\pm 0.75$ $\mu$m}, i.e. a value of the same order of magnitude of the interface thickness in our experiments, marking the transition between the nearly-exponential decay of the tension and the square gradient scaling. Moreover, $l_D$ is lower than $L$, in analogy to previous findings for colloidal and polymeric fluids \cite{truzzolilloNonequilibriumInterfacialTension2016c}.
%In this respect it has been argued \cite{orzaPatternsLongRange1997,guazzelliFluctuationsInstabilitySedimentation2011,katsuragiJammingGrowthDynamical2010,wuEnhancedConcentrationFluctuations1991} that the emergence of a characteristic length larger than typical structural length scales is a consequence of a system being strongly driven as observed in a variety of systems, from flowing granular matter, to colloidal and polymer systems. 

\rv{We find} therefore the existence of two distinct decay regimes for interfacial stresses at miscible liquid/liquid boundaries: at short time scales, \rv{in the presence of }sharp gradients, high-order terms in the free-energy gradient expansion dominate over the square gradient term. At large \rv{$t_c$}, the concentration gradients are small and the SG term dominates, \rv{accounting for $\Gamma_e$ as obtained by spinning drop tensiometry at large $t_c$ (star in Fig \ref{fig4}-b).}
%: its contribution is necessary to account for the point that we obtained analyzing spinning drop deformations at large $t_c$ (star in Fig \ref{fig4}-b). 

\rv{We further used the best fit through Eq. \ref{efftens_tot}) and }the experimental $t_c$ to \rv{estimate} the effective tension for those rates of glycerol for which \rv{no \textit{C}-branch could be observed, finding $\Gamma_e=$ 1.91 $\mu$N/m and 0.96 $\mu$N/m} for $Q_{Gly}=6$ and 3 $\mu$/min, respectively. We used these values to compute the prefactor $\alpha$ \rv{in} Eq.\ref{capillary} and report \rv{in Fig. \ref{fig4}} the wavenumber as a function of the rescaled frequency $\omega/\alpha$ also for the two lowest $Q_{G}$ (green symbols). Remarkably, we find that the wavenumber does not show any clear dependence over almost 2 decades in frequency (full range shown in \cite{SeeSupplementalMaterial}), \rv{fluctuating} around an average value $k_{D}=18\pm2$ mm$^{-1}$, and merging to the highest wavenumber region of the capillary branch, akin to what observed in \cite{huViscousWaveBreaking2018} for silicone oil and ethanol, a pair of immiscible fluids with a viscosity contrast similar to glycerol and water (see \cite{SeeSupplementalMaterial} for a direct comparison). 

\rv{Here therefore, the same two branches observed in \cite{huViscousWaveBreaking2018} for a pair of immiscible fluids are also seen for two inter-diffusing liquids, in contrast to the results reported in \cite{huViscousWaveBreaking2018}, where the \textit{C}- and the \textit{CD}- branches were absent for miscible silicone oils.} %thus suggest two different mechanisms driving wave propagation at the boundary between miscible fluids, giving rise to two distinct branches of the wave dispersion. One of them is characterized by the same inertial scaling, the \textit{D}-branch here, observed in \cite{huViscousWaveBreaking2018}, while the other (the \textit{C}-branch) and the apparent transition between them (the \textit{CD} regime)  are observed in the present work and are absent in \cite{huViscousWaveBreaking2018}. 
This discrepancy requires further inspection. 

\rv{Our} experiments have been performed in a channel ($W=1$mm) wider than that of \cite{huViscousWaveBreaking2018} ($W=0.25$mm). We therefore start inquiring whether confinement is key for the appearance of waves following the inertial scaling $k\sim\omega^0$, \rv{i.e.}  whether confinement hampers the capillary regime $k\sim\omega^{2/3}$. We have performed further co-flow experiments employing a narrower channel, $W=$0.25 mm, in the same range of Reynolds numbers (\cite{SeeSupplementalMaterial}). Notably, we did not observe the onset of a capillary branch in the dispersion relation, \rv{in excellent qualitative and quantitative agreement with} the results of \cite{huViscousWaveBreaking2018} for miscible fluids (Fig. \ref{fig4}-c).
A further clue of the key role played by confinement is given by the fact that the transition from \textit{C} to \textit{CD} waves occurs \rv{as the interface position $Y$ approaches the wave amplitude~\cite{SeeSupplementalMaterial}, with the latter sharply decreasing by further approaching the wall at $y=0$.}   
We \rv{thus} conclude that lateral confinement is responsible for suppressing capillary waves under weak tensions and we propose that the mode selection in this case is driven by a minimum-dissipation principle dictating the minimum wavelength emerging for this instability, as \rv{reported} for other instabilities where capillary stresses at play are very low \cite{patersonFingeringMiscibleFluids1985a}. It is not surprising therefore that, what has been called inertial regime, defines the upper \rv{(large-$k$)} bound of the unstable region in Fig. \ref{fig4}-a and for the experimental results reported in \cite{huViscousWaveBreaking2018}. \rv{Having established the existence of capillary and dissipative} waves in confined co-flows, the \rv{intermediate} \textit{CD}-regime observed for $Q_{G}\geq$ 8 $\mu$l/l %that is characterized by the departure from pure capillary (\textit{C}) branch towards the inertial one, 
can be ascribed to the competition between the \rv{\textit{C} and \textit{D} modes}. We thus consider that \rv{the \textit{CD} modes correspond to} a transition region in a very general stability diagram, sketched in Fig. \ref{fig4}-d. As a final remark, we emphasize that the same \rv{\textit{C} and \textit{D}} branches delimiting the instability region of the diagram appear in both miscible and immiscible fluids, \rv{see }Fig. SM7 in \cite{SeeSupplementalMaterial}. \rv{This }overlap of the wave dispersions observed in pairs of immiscible and fully miscible fluids further supports the importance \rv{and generality} of interfacial stresses at diffuse interfaces far from equilibrium.    

%In this letter, we reported on the existence of capillary waves in miscible fluids. 
\rv{To summarize, we have shown that capillary waves, hitherto considered to be a distinctive phenomenon of interfaces between immiscible fluids, emerge also at diffuse boundaries between fully miscible liquids. Their wave propagation is dominated by a transient interfacial tension for high flow rates and large concentration gradients}, while dissipation and confinement dominate in the inertial regime, where one single wavenumber emerges \rv{and capillary waves are suppressed}. 
%we reported the existence of waves that emerge spontaneously at the interface between miscible liquids in viscosity-stratified microfluidic flows, following two types of dispersive behavior. Wave propagation is dominated respectively by interfacial stresses for high flow rates and large concentration gradients (\textit{C}-branch), 
The transition from capillary waves to the inertial regime is generally smooth and occurs when the rate of the fastest fluid is decreased so that diffusion smears out interfaces, interfacial stresses become vanishingly small, and waves get concomitantly more confined. 
We found very good agreement between the measured values of the effective tension and a model \cite{truzzolilloNonequilibriumInterfacialTension2016c} that goes beyond the well-known second order gradient expansion of pressures and mixing free energy in systems far from equilibrium \cite{kortewegFormeQuePrennent1901,rousarCONTINUUMANALYSISSURFACE1994}. Our results indicate that interfacial tension in miscible molecular fluids \rvv{decays} much faster right after contact than the \rv{$t^{-1/2}$ scaling predicted by square gradient models, as hinted at} by other results \rvv{based on the detection of capillary waves in miscible near-critical fluids} in static conditions \cite{cicutaCapillarytobulkCrossoverNonequilibrium2001}, while the square gradient approximation holds at later stages \cite{carbonaroUltralowEffectiveInterfacial2020a}.
\rv{The capillary wave patterns reported here for} microfluidic flows of miscible liquids are a novel and unexpected phenomenon, that we propose as a \rv{measurement strategy for accessing transient interfacial tension down to the millisecond time scale.} \rv{Our work paves the way for a thorough understanding of capillary phenomena and flow stability in miscible fluids, which play a key role in a wide range
of research fields of both academic and practical interest,
from non-equilibrium thermodynamics \cite{mauriNonEquilibriumThermodynamicsMultiphase2013} and geosciences \cite{morraRoleKortewegStresses2008} to oil recovery \cite{babadagliDevelopmentMatureOil2007}, filtration and flow in porous media (e.g., in a chromatography column \cite{swernathEffectKortewegStress2010}), fluid removal \cite{petitjeansMiscibleDisplacementsCapillary1996}, and self-propulsion of droplets and colloids of biological interest \cite{banSelfgeneratedMotionDroplets2010,banSelfPropelledVesiclesInduced2016}.} 
 
\begin{acknowledgments}
DT warmly thanks Prof. Laura Casanellas for help in manufacturing the microchannels. LC gratefully acknowledges support from the Institut Universitaire de France (IUF). We gratefully acknowledge support from the Centre national d’\'{e}tudes spatiales (CNES).
\rv{RG acknowledges support of the Department of Atomic Energy, Government of India, under the project no RTI4001.}
\end{acknowledgments}

\appendix
%\section{Appendixes}
%\bibliography{coflow_resub}% Produces the bibliography via BibTeX.

\begin{thebibliography}{33}%
\makeatletter
\providecommand \@ifxundefined [1]{%
 \@ifx{#1\undefined}
}%
\providecommand \@ifnum [1]{%
 \ifnum #1\expandafter \@firstoftwo
 \else \expandafter \@secondoftwo
 \fi
}%
\providecommand \@ifx [1]{%
 \ifx #1\expandafter \@firstoftwo
 \else \expandafter \@secondoftwo
 \fi
}%
\providecommand \natexlab [1]{#1}%
\providecommand \enquote  [1]{``#1''}%
\providecommand \bibnamefont  [1]{#1}%
\providecommand \bibfnamefont [1]{#1}%
\providecommand \citenamefont [1]{#1}%
\providecommand \href@noop [0]{\@secondoftwo}%
\providecommand \href [0]{\begingroup \@sanitize@url \@href}%
\providecommand \@href[1]{\@@startlink{#1}\@@href}%
\providecommand \@@href[1]{\endgroup#1\@@endlink}%
\providecommand \@sanitize@url [0]{\catcode `\\12\catcode `\$12\catcode
  `\&12\catcode `\#12\catcode `\^12\catcode `\_12\catcode `\%12\relax}%
\providecommand \@@startlink[1]{}%
\providecommand \@@endlink[0]{}%
\providecommand \url  [0]{\begingroup\@sanitize@url \@url }%
\providecommand \@url [1]{\endgroup\@href {#1}{\urlprefix }}%
\providecommand \urlprefix  [0]{URL }%
\providecommand \Eprint [0]{\href }%
\providecommand \doibase [0]{https://doi.org/}%
\providecommand \selectlanguage [0]{\@gobble}%
\providecommand \bibinfo  [0]{\@secondoftwo}%
\providecommand \bibfield  [0]{\@secondoftwo}%
\providecommand \translation [1]{[#1]}%
\providecommand \BibitemOpen [0]{}%
\providecommand \bibitemStop [0]{}%
\providecommand \bibitemNoStop [0]{.\EOS\space}%
\providecommand \EOS [0]{\spacefactor3000\relax}%
\providecommand \BibitemShut  [1]{\csname bibitem#1\endcsname}%
\let\auto@bib@innerbib\@empty
%</preamble>
\bibitem [{\citenamefont {Lamb}(1932)}]{lambHydrodynamics1932}%
  \BibitemOpen
  \bibfield  {author} {\bibinfo {author} {\bibfnamefont {H.}~\bibnamefont
  {Lamb}},\ }\href@noop {} {\emph {\bibinfo {title} {Hydrodynamics}}},\
  \bibinfo {edition} {6th}\ ed.\ (\bibinfo  {publisher} {Cambridge University
  Press.},\ \bibinfo {address} {New York},\ \bibinfo {year} {1932})\BibitemShut
  {NoStop}%
\bibitem [{\citenamefont {{Rayleigh}}(1877)}]{rayleighProgressiveWaves1877}%
  \BibitemOpen
  \bibfield  {author} {\bibinfo {author} {\bibnamefont {{Rayleigh}}},\
  }\bibfield  {title} {\bibinfo {title} {On {{Progressive Waves}}},\ }\href
  {https://doi.org/10.1112/plms/s1-9.1.21} {\bibfield  {journal} {\bibinfo
  {journal} {Proceedings of the London Mathematical Society}\ }\textbf
  {\bibinfo {volume} {s1-9}},\ \bibinfo {pages} {21} (\bibinfo {year}
  {1877})}\BibitemShut {NoStop}%
\bibitem [{\citenamefont {Chou}\ \emph {et~al.}(1995)\citenamefont {Chou},
  \citenamefont {Lucas},\ and\ \citenamefont
  {Stone}}]{chouCapillaryWaveScattering1995}%
  \BibitemOpen
  \bibfield  {author} {\bibinfo {author} {\bibfnamefont {T.}~\bibnamefont
  {Chou}}, \bibinfo {author} {\bibfnamefont {S.~K.}\ \bibnamefont {Lucas}},\
  and\ \bibinfo {author} {\bibfnamefont {H.~A.}\ \bibnamefont {Stone}},\
  }\bibfield  {title} {\bibinfo {title} {Capillary wave scattering from a
  surfactant domain},\ }\href {https://doi.org/10.1063/1.868502} {\bibfield
  {journal} {\bibinfo  {journal} {Physics of Fluids}\ }\textbf {\bibinfo
  {volume} {7}},\ \bibinfo {pages} {1872} (\bibinfo {year} {1995})}\BibitemShut
  {NoStop}%
\bibitem [{\citenamefont {Bobb}\ \emph {et~al.}(1979)\citenamefont {Bobb},
  \citenamefont {Ferguson},\ and\ \citenamefont
  {Rankin}}]{bobbCapillaryWaveMeasurements1979}%
  \BibitemOpen
  \bibfield  {author} {\bibinfo {author} {\bibfnamefont {L.~C.}\ \bibnamefont
  {Bobb}}, \bibinfo {author} {\bibfnamefont {G.}~\bibnamefont {Ferguson}},\
  and\ \bibinfo {author} {\bibfnamefont {M.}~\bibnamefont {Rankin}},\
  }\bibfield  {title} {\bibinfo {title} {Capillary wave measurements},\ }\href
  {https://doi.org/10.1364/AO.18.001167} {\bibfield  {journal} {\bibinfo
  {journal} {Appl. Opt.}\ }\textbf {\bibinfo {volume} {18}},\ \bibinfo {pages}
  {1167} (\bibinfo {year} {1979})}\BibitemShut {NoStop}%
\bibitem [{\citenamefont {Behroozi}\ \emph {et~al.}(2001)\citenamefont
  {Behroozi}, \citenamefont {Lambert},\ and\ \citenamefont
  {Buhrow}}]{behrooziDirectMeasurementAttenuation2001}%
  \BibitemOpen
  \bibfield  {author} {\bibinfo {author} {\bibfnamefont {F.}~\bibnamefont
  {Behroozi}}, \bibinfo {author} {\bibfnamefont {B.}~\bibnamefont {Lambert}},\
  and\ \bibinfo {author} {\bibfnamefont {B.}~\bibnamefont {Buhrow}},\
  }\bibfield  {title} {\bibinfo {title} {Direct measurement of the attenuation
  of capillary waves by laser interferometry: {{Noncontact}} determination of
  viscosity},\ }\href {https://doi.org/10.1063/1.1365413} {\bibfield  {journal}
  {\bibinfo  {journal} {Appl. Phys. Lett.}\ }\textbf {\bibinfo {volume} {78}},\
  \bibinfo {pages} {2399} (\bibinfo {year} {2001})}\BibitemShut {NoStop}%
\bibitem [{\citenamefont {Sohl}\ \emph {et~al.}(1978)\citenamefont {Sohl},
  \citenamefont {Miyano},\ and\ \citenamefont
  {Ketterson}}]{sohlNovelTechniqueDynamic1978}%
  \BibitemOpen
  \bibfield  {author} {\bibinfo {author} {\bibfnamefont {C.~H.}\ \bibnamefont
  {Sohl}}, \bibinfo {author} {\bibfnamefont {K.}~\bibnamefont {Miyano}},\ and\
  \bibinfo {author} {\bibfnamefont {J.~B.}\ \bibnamefont {Ketterson}},\
  }\bibfield  {title} {\bibinfo {title} {Novel technique for dynamic surface
  tension and viscosity measurements at liquid--gas interfaces},\ }\href
  {https://doi.org/10.1063/1.1135288} {\bibfield  {journal} {\bibinfo
  {journal} {Review of Scientific Instruments}\ }\textbf {\bibinfo {volume}
  {49}},\ \bibinfo {pages} {1464} (\bibinfo {year} {1978})}\BibitemShut
  {NoStop}%
\bibitem [{\citenamefont {Hu}\ and\ \citenamefont
  {Cubaud}(2018)}]{huViscousWaveBreaking2018}%
  \BibitemOpen
  \bibfield  {author} {\bibinfo {author} {\bibfnamefont {X.}~\bibnamefont
  {Hu}}\ and\ \bibinfo {author} {\bibfnamefont {T.}~\bibnamefont {Cubaud}},\
  }\bibfield  {title} {\bibinfo {title} {Viscous {{Wave Breaking}} and
  {{Ligament Formation}} in {{Microfluidic Systems}}},\ }\href
  {https://doi.org/10.1103/PhysRevLett.121.044502} {\bibfield  {journal}
  {\bibinfo  {journal} {Phys. Rev. Lett.}\ }\textbf {\bibinfo {volume} {121}},\
  \bibinfo {pages} {044502} (\bibinfo {year} {2018})}\BibitemShut {NoStop}%
\bibitem [{\citenamefont
  {Bier}(2015)}]{bierNonequilibriumInterfacialTension2015}%
  \BibitemOpen
  \bibfield  {author} {\bibinfo {author} {\bibfnamefont {M.}~\bibnamefont
  {Bier}},\ }\bibfield  {title} {\bibinfo {title} {Nonequilibrium interfacial
  tension during relaxation},\ }\href
  {https://doi.org/10.1103/PhysRevE.92.042128} {\bibfield  {journal} {\bibinfo
  {journal} {Phys. Rev. E}\ }\textbf {\bibinfo {volume} {92}},\ \bibinfo
  {pages} {042128} (\bibinfo {year} {2015})}\BibitemShut {NoStop}%
\bibitem [{\citenamefont {Truzzolillo}\ \emph {et~al.}(2016)\citenamefont
  {Truzzolillo}, \citenamefont {Mora}, \citenamefont {Dupas},\ and\
  \citenamefont
  {Cipelletti}}]{truzzolilloNonequilibriumInterfacialTension2016c}%
  \BibitemOpen
  \bibfield  {author} {\bibinfo {author} {\bibfnamefont {D.}~\bibnamefont
  {Truzzolillo}}, \bibinfo {author} {\bibfnamefont {S.}~\bibnamefont {Mora}},
  \bibinfo {author} {\bibfnamefont {C.}~\bibnamefont {Dupas}},\ and\ \bibinfo
  {author} {\bibfnamefont {L.}~\bibnamefont {Cipelletti}},\ }\bibfield  {title}
  {\bibinfo {title} {Nonequilibrium {{Interfacial Tension}} in {{Simple}} and
  {{Complex Fluids}}},\ }\href {https://doi.org/10.1103/PhysRevX.6.041057}
  {\bibfield  {journal} {\bibinfo  {journal} {Phys. Rev. X}\ }\textbf {\bibinfo
  {volume} {6}},\ \bibinfo {pages} {041057} (\bibinfo {year}
  {2016})}\BibitemShut {NoStop}%
\bibitem [{\citenamefont {Suzuki}\ \emph {et~al.}(2020)\citenamefont {Suzuki},
  \citenamefont {Quah}, \citenamefont {Ban}, \citenamefont {Mishra},\ and\
  \citenamefont {Nagatsu}}]{suzukiExperimentalStudyMiscible2020}%
  \BibitemOpen
  \bibfield  {author} {\bibinfo {author} {\bibfnamefont {R.~X.}\ \bibnamefont
  {Suzuki}}, \bibinfo {author} {\bibfnamefont {F.~W.}\ \bibnamefont {Quah}},
  \bibinfo {author} {\bibfnamefont {T.}~\bibnamefont {Ban}}, \bibinfo {author}
  {\bibfnamefont {M.}~\bibnamefont {Mishra}},\ and\ \bibinfo {author}
  {\bibfnamefont {Y.}~\bibnamefont {Nagatsu}},\ }\bibfield  {title} {\bibinfo
  {title} {Experimental study of miscible viscous fingering with different
  effective interfacial tension},\ }\href {https://doi.org/10.1063/5.0030152}
  {\bibfield  {journal} {\bibinfo  {journal} {AIP Advances}\ }\textbf {\bibinfo
  {volume} {10}},\ \bibinfo {pages} {115219} (\bibinfo {year}
  {2020})}\BibitemShut {NoStop}%
\bibitem [{\citenamefont {Gowda}\ \emph {et~al.}(2021)\citenamefont {Gowda},
  \citenamefont {Rydefalk}, \citenamefont {S{\"o}derberg},\ and\ \citenamefont
  {Lundell}}]{gowdaFormationColloidalThreads2021}%
  \BibitemOpen
  \bibfield  {author} {\bibinfo {author} {\bibfnamefont {V.~K.}\ \bibnamefont
  {Gowda}}, \bibinfo {author} {\bibfnamefont {C.}~\bibnamefont {Rydefalk}},
  \bibinfo {author} {\bibfnamefont {L.~D.}\ \bibnamefont {S{\"o}derberg}},\
  and\ \bibinfo {author} {\bibfnamefont {F.}~\bibnamefont {Lundell}},\
  }\bibfield  {title} {\bibinfo {title} {Formation of colloidal threads in
  geometrically varying flow-focusing channels},\ }\href
  {https://doi.org/10.1103/PhysRevFluids.6.114001} {\bibfield  {journal}
  {\bibinfo  {journal} {Phys. Rev. Fluids}\ }\textbf {\bibinfo {volume} {6}},\
  \bibinfo {pages} {114001} (\bibinfo {year} {2021})}\BibitemShut {NoStop}%
\bibitem [{\citenamefont {Vorobev}\ \emph {et~al.}(2021)\citenamefont
  {Vorobev}, \citenamefont {Prokopev},\ and\ \citenamefont
  {Lyubimova}}]{vorobevNonequilibriumCapillaryPressure2021}%
  \BibitemOpen
  \bibfield  {author} {\bibinfo {author} {\bibfnamefont {A.}~\bibnamefont
  {Vorobev}}, \bibinfo {author} {\bibfnamefont {S.}~\bibnamefont {Prokopev}},\
  and\ \bibinfo {author} {\bibfnamefont {T.}~\bibnamefont {Lyubimova}},\
  }\bibfield  {title} {\bibinfo {title} {Nonequilibrium {{Capillary Pressure}}
  of a {{Miscible Meniscus}}},\ }\href
  {https://doi.org/10.1021/acs.langmuir.0c03633} {\bibfield  {journal}
  {\bibinfo  {journal} {Langmuir}\ }\textbf {\bibinfo {volume} {37}},\ \bibinfo
  {pages} {4817} (\bibinfo {year} {2021})}\BibitemShut {NoStop}%
\bibitem [{\citenamefont {Lacaze}\ \emph {et~al.}(2010)\citenamefont {Lacaze},
  \citenamefont {Guenoun}, \citenamefont {Beysens}, \citenamefont {Delsanti},
  \citenamefont {Petitjeans},\ and\ \citenamefont
  {Kurowski}}]{lacazeTransientSurfaceTension2010}%
  \BibitemOpen
  \bibfield  {author} {\bibinfo {author} {\bibfnamefont {L.}~\bibnamefont
  {Lacaze}}, \bibinfo {author} {\bibfnamefont {P.}~\bibnamefont {Guenoun}},
  \bibinfo {author} {\bibfnamefont {D.~N.}\ \bibnamefont {Beysens}}, \bibinfo
  {author} {\bibfnamefont {M.}~\bibnamefont {Delsanti}}, \bibinfo {author}
  {\bibfnamefont {P.}~\bibnamefont {Petitjeans}},\ and\ \bibinfo {author}
  {\bibfnamefont {P.}~\bibnamefont {Kurowski}},\ }\bibfield  {title} {\bibinfo
  {title} {Transient surface tension in miscible liquids},\ }\href
  {https://doi.org/10.1103/PhysRevE.82.041606} {\bibfield  {journal} {\bibinfo
  {journal} {Physical Review E}\ }\textbf {\bibinfo {volume} {82}},\ \bibinfo
  {pages} {041606} (\bibinfo {year} {2010})}\BibitemShut {NoStop}%
\bibitem [{\citenamefont {Korteweg}(1901)}]{kortewegFormeQuePrennent1901}%
  \BibitemOpen
  \bibfield  {author} {\bibinfo {author} {\bibfnamefont {D.}~\bibnamefont
  {Korteweg}},\ }\bibfield  {title} {\bibinfo {title} {Sur la forme que
  prennent les {\'e}quations du mouvements des fluides si l'on tient compte des
  forces capillaires caus{\'e}es par des variations de densit{\'e}
  consid{\'e}rables mais connues et sur la th{\'e}orie de la capillarit{\'e}
  dans l'hypoth{\`e}se d'une variation continue de la densit{\'e}},\
  }\href@noop {} {\bibfield  {journal} {\bibinfo  {journal} {Arch. Neerland.
  Sci. Exact. et Naturell.}\ }\textbf {\bibinfo {volume} {6}},\ \bibinfo
  {pages} {1} (\bibinfo {year} {1901})}\BibitemShut {NoStop}%
\bibitem [{\citenamefont {Pojman}\ \emph {et~al.}(2006)\citenamefont {Pojman},
  \citenamefont {Whitmore}, \citenamefont {Turco~Liveri}, \citenamefont
  {Lombardo}, \citenamefont {Marszalek}, \citenamefont {Parker},\ and\
  \citenamefont {Zoltowski}}]{pojmanEvidenceExistenceEffective2006b}%
  \BibitemOpen
  \bibfield  {author} {\bibinfo {author} {\bibfnamefont {J.~A.}\ \bibnamefont
  {Pojman}}, \bibinfo {author} {\bibfnamefont {C.}~\bibnamefont {Whitmore}},
  \bibinfo {author} {\bibfnamefont {M.~L.}\ \bibnamefont {Turco~Liveri}},
  \bibinfo {author} {\bibfnamefont {R.}~\bibnamefont {Lombardo}}, \bibinfo
  {author} {\bibfnamefont {J.}~\bibnamefont {Marszalek}}, \bibinfo {author}
  {\bibfnamefont {R.}~\bibnamefont {Parker}},\ and\ \bibinfo {author}
  {\bibfnamefont {B.}~\bibnamefont {Zoltowski}},\ }\bibfield  {title} {\bibinfo
  {title} {Evidence for the {{Existence}} of an {{Effective Interfacial
  Tension}} between {{Miscible Fluids}}: {{Isobutyric Acid}}-{{Water}} and
  1-{{Butanol}}-{{Water}} in a {{Spinning-Drop Tensiometer}}},\ }\href
  {https://doi.org/10.1021/la052111n} {\bibfield  {journal} {\bibinfo
  {journal} {Langmuir}\ }\textbf {\bibinfo {volume} {22}},\ \bibinfo {pages}
  {2569} (\bibinfo {year} {2006})}\BibitemShut {NoStop}%
\bibitem [{\citenamefont {Cicuta}\ \emph {et~al.}(2001)\citenamefont {Cicuta},
  \citenamefont {Vailati},\ and\ \citenamefont
  {Giglio}}]{cicutaCapillarytobulkCrossoverNonequilibrium2001}%
  \BibitemOpen
  \bibfield  {author} {\bibinfo {author} {\bibfnamefont {P.}~\bibnamefont
  {Cicuta}}, \bibinfo {author} {\bibfnamefont {A.}~\bibnamefont {Vailati}},\
  and\ \bibinfo {author} {\bibfnamefont {M.}~\bibnamefont {Giglio}},\
  }\bibfield  {title} {\bibinfo {title} {Capillary-to-bulk crossover of
  nonequilibrium fluctuations in the free diffusion of a near-critical binary
  liquid mixture},\ }\href {https://doi.org/10.1364/AO.40.004140} {\bibfield
  {journal} {\bibinfo  {journal} {Applied Optics}\ }\textbf {\bibinfo {volume}
  {40}},\ \bibinfo {pages} {4140} (\bibinfo {year} {2001})}\BibitemShut
  {NoStop}%
\bibitem [{\citenamefont {Smith}\ \emph {et~al.}(1981)\citenamefont {Smith},
  \citenamefont {Van De~Ven},\ and\ \citenamefont
  {Mason}}]{smithTransientInterfacialTension1981}%
  \BibitemOpen
  \bibfield  {author} {\bibinfo {author} {\bibfnamefont {P.}~\bibnamefont
  {Smith}}, \bibinfo {author} {\bibfnamefont {T.}~\bibnamefont {Van De~Ven}},\
  and\ \bibinfo {author} {\bibfnamefont {S.}~\bibnamefont {Mason}},\ }\bibfield
   {title} {\bibinfo {title} {The transient interfacial tension between two
  miscible fluids},\ }\href {https://doi.org/10.1016/0021-9797(81)90186-7}
  {\bibfield  {journal} {\bibinfo  {journal} {Journal of Colloid and Interface
  Science}\ }\textbf {\bibinfo {volume} {80}},\ \bibinfo {pages} {302}
  (\bibinfo {year} {1981})}\BibitemShut {NoStop}%
\bibitem [{\citenamefont {Truzzolillo}\ \emph {et~al.}(2014)\citenamefont
  {Truzzolillo}, \citenamefont {Mora}, \citenamefont {Dupas},\ and\
  \citenamefont {Cipelletti}}]{truzzolilloEquilibriumSurfaceTension2014}%
  \BibitemOpen
  \bibfield  {author} {\bibinfo {author} {\bibfnamefont {D.}~\bibnamefont
  {Truzzolillo}}, \bibinfo {author} {\bibfnamefont {S.}~\bibnamefont {Mora}},
  \bibinfo {author} {\bibfnamefont {C.}~\bibnamefont {Dupas}},\ and\ \bibinfo
  {author} {\bibfnamefont {L.}~\bibnamefont {Cipelletti}},\ }\bibfield  {title}
  {\bibinfo {title} {Off-{{Equilibrium Surface Tension}} in {{Colloidal
  Suspensions}}},\ }\href {https://doi.org/10.1103/PhysRevLett.112.128303}
  {\bibfield  {journal} {\bibinfo  {journal} {Physical Review Letters}\
  }\textbf {\bibinfo {volume} {112}},\ \bibinfo {pages} {128303} (\bibinfo
  {year} {2014})}\BibitemShut {NoStop}%
\bibitem [{See()}]{SeeSupplementalMaterial}%
  \BibitemOpen
  \href@noop {} {\bibinfo {title} {See {{Supplemental Material}} at [{{URL}}
  will be inserted by publisher] for: {{I}}) {{Fabrication}} of the
  microchannels; {{II}}) {{Visualization}} and tracking of the waves; {{III}})
  {{Low-Reynolds-number}} instabilities; {{IV}}) {{Base}} flow profile of two
  co-flowing fluids in a rectangular channel; {{V}}) {{Accuracy}} of the
  interface position and interfacial velocity of the unperturbed interface;
  {{VI}}) {{Wave}} amplitude, confinement and critical rates of water {{VII}})
  {{Stability}} diagram; {{VIII}}) {{Calculation}} of {{$\Gamma_{SG}(t_c)$}}
  for water-glycerol interfaces; {{IX}}) {{Proxies}} and time dependence of the
  interface thickness $\delta(t_c)$; {{X}}) {{Best}} fitting values for {{$L$}}
  and {{$\varepsilon$}} and comparison with previous literature data; {{XI}})
  {{Comparison}} between miscible and immiscible fluids ({{Data}} from
  [7]). \rvv{{{Supplemental Material}} includes refs \cite{guillotViscosimeterMicrofluidicChip2006,cubaudHighviscosityFluidThreads2009,demenechTransitionSqueezingDripping2008,tabelingIntroductionMicrofluidics2023,friendFabricationMicrofluidicDevices2010,govindarajanInstabilitiesViscosityStratifiedFlow2014,govindarajanEffectMiscibilityLinear2004,tosunCriticalReynoldsNumber1988,sahuLinearStabilityAnalysis2016,stilesHydrodynamicControlInterface2004,stoneEngineeringFlowsSmall2004,derricoDiffusionCoefficientsBinary2004a,kjelstrupNonequilibriumThermodynamicsHeterogeneous2008,orzaPatternsLongRange1997,guazzelliFluctuationsInstabilitySedimentation2011,katsuragiJammingGrowthDynamical2010,wuEnhancedConcentrationFluctuations1991}} }}\BibitemShut {NoStop}%
\bibitem [{\citenamefont {Landau}\ and\ \citenamefont
  {Lifshitz}(1987)}]{landauCourseTheoreticalPhysics1987}%
  \BibitemOpen
  \bibfield  {author} {\bibinfo {author} {\bibfnamefont {L.~D.}\ \bibnamefont
  {Landau}}\ and\ \bibinfo {author} {\bibfnamefont {E.~M.}\ \bibnamefont
  {Lifshitz}},\ }\href@noop {} {\emph {\bibinfo {title} {Course of Theoretical
  Physics, {{Volume}} 6 - {{Fluid Mechanics}}.}}}\ (\bibinfo  {publisher}
  {Pergamon Press},\ \bibinfo {year} {1987})\BibitemShut {NoStop}%
\bibitem [{\citenamefont {Denner}\ and\ \citenamefont
  {Van~Wachem}(2015)}]{dennerNumericalTimestepRestrictions2015}%
  \BibitemOpen
  \bibfield  {author} {\bibinfo {author} {\bibfnamefont {F.}~\bibnamefont
  {Denner}}\ and\ \bibinfo {author} {\bibfnamefont {B.~G.}\ \bibnamefont
  {Van~Wachem}},\ }\bibfield  {title} {\bibinfo {title} {Numerical time-step
  restrictions as a result of capillary waves},\ }\href
  {https://doi.org/10.1016/j.jcp.2015.01.021} {\bibfield  {journal} {\bibinfo
  {journal} {Journal of Computational Physics}\ }\textbf {\bibinfo {volume}
  {285}},\ \bibinfo {pages} {24} (\bibinfo {year} {2015})}\BibitemShut
  {NoStop}%
\bibitem [{\citenamefont {Giamagas}\ \emph {et~al.}(2023)\citenamefont
  {Giamagas}, \citenamefont {Zonta}, \citenamefont {Roccon},\ and\
  \citenamefont {Soldati}}]{giamagasPropagationCapillaryWaves2023}%
  \BibitemOpen
  \bibfield  {author} {\bibinfo {author} {\bibfnamefont {G.}~\bibnamefont
  {Giamagas}}, \bibinfo {author} {\bibfnamefont {F.}~\bibnamefont {Zonta}},
  \bibinfo {author} {\bibfnamefont {A.}~\bibnamefont {Roccon}},\ and\ \bibinfo
  {author} {\bibfnamefont {A.}~\bibnamefont {Soldati}},\ }\bibfield  {title}
  {\bibinfo {title} {Propagation of capillary waves in two-layer oil--water
  turbulent flow},\ }\href {https://doi.org/10.1017/jfm.2023.189} {\bibfield
  {journal} {\bibinfo  {journal} {J. Fluid Mech.}\ }\textbf {\bibinfo {volume}
  {960}},\ \bibinfo {pages} {A5} (\bibinfo {year} {2023})}\BibitemShut
  {NoStop}%
\bibitem [{\citenamefont {Carbonaro}\ \emph {et~al.}(2020)\citenamefont
  {Carbonaro}, \citenamefont {Cipelletti},\ and\ \citenamefont
  {Truzzolillo}}]{carbonaroUltralowEffectiveInterfacial2020a}%
  \BibitemOpen
  \bibfield  {author} {\bibinfo {author} {\bibfnamefont {A.}~\bibnamefont
  {Carbonaro}}, \bibinfo {author} {\bibfnamefont {L.}~\bibnamefont
  {Cipelletti}},\ and\ \bibinfo {author} {\bibfnamefont {D.}~\bibnamefont
  {Truzzolillo}},\ }\bibfield  {title} {\bibinfo {title} {Ultralow effective
  interfacial tension between miscible molecular fluids},\ }\href
  {https://doi.org/10.1103/PhysRevFluids.5.074001} {\bibfield  {journal}
  {\bibinfo  {journal} {Phys. Rev. Fluids}\ }\textbf {\bibinfo {volume} {5}},\
  \bibinfo {pages} {074001} (\bibinfo {year} {2020})}\BibitemShut {NoStop}%
\bibitem [{\citenamefont
  {Petitjeans}(1996)}]{petitjeansPetitjeanTensionSurface1996a}%
  \BibitemOpen
  \bibfield  {author} {\bibinfo {author} {\bibfnamefont {P.}~\bibnamefont
  {Petitjeans}},\ }\bibfield  {title} {\bibinfo {title} {{{Une}}
  tension de surface pour le fluides miscibles},\ }\href@noop {} {\bibfield
  {journal} {\bibinfo  {journal} {C.R. Acad Sci. Paris}\ ,\ \bibinfo {pages}
  {673}} (\bibinfo {year} {1996})}\BibitemShut {NoStop}%
\bibitem [{\citenamefont
  {Paterson}(1985)}]{patersonFingeringMiscibleFluids1985a}%
  \BibitemOpen
  \bibfield  {author} {\bibinfo {author} {\bibfnamefont {L.}~\bibnamefont
  {Paterson}},\ }\bibfield  {title} {\bibinfo {title} {Fingering with miscible
  fluids in a {{Hele Shaw}} cell},\ }\href {https://doi.org/10.1063/1.865195}
  {\bibfield  {journal} {\bibinfo  {journal} {Physics of Fluids}\ }\textbf
  {\bibinfo {volume} {28}},\ \bibinfo {pages} {26} (\bibinfo {year}
  {1985})}\BibitemShut {NoStop}%
\bibitem [{\citenamefont {Rou{\v s}ar}\ and\ \citenamefont
  {Nauman}(1994)}]{rousarCONTINUUMANALYSISSURFACE1994}%
  \BibitemOpen
  \bibfield  {author} {\bibinfo {author} {\bibfnamefont {I.}~\bibnamefont
  {Rou{\v s}ar}}\ and\ \bibinfo {author} {\bibfnamefont {E.}~\bibnamefont
  {Nauman}},\ }\bibfield  {title} {\bibinfo {title} {A continuum analysis of
  surface tension in nonequilibrium systems},\ }\href
  {https://doi.org/10.1080/00986449408936247} {\bibfield  {journal} {\bibinfo
  {journal} {Chemical Engineering Communications}\ }\textbf {\bibinfo {volume}
  {129}},\ \bibinfo {pages} {19} (\bibinfo {year} {1994})}\BibitemShut
  {NoStop}%
\bibitem [{\citenamefont
  {Mauri}(2013)}]{mauriNonEquilibriumThermodynamicsMultiphase2013}%
  \BibitemOpen
  \bibfield  {author} {\bibinfo {author} {\bibfnamefont {R.}~\bibnamefont
  {Mauri}},\ }\href {https://doi.org/10.1007/978-94-007-5461-4} {\emph
  {\bibinfo {title} {Non-{{Equilibrium Thermodynamics}} in {{Multiphase
  Flows}}}}},\ Soft and {{Biological Matter}}\ (\bibinfo  {publisher} {Springer
  Netherlands},\ \bibinfo {address} {Dordrecht},\ \bibinfo {year}
  {2013})\BibitemShut {NoStop}%
\bibitem [{\citenamefont {Morra}\ and\ \citenamefont
  {Yuen}(2008)}]{morraRoleKortewegStresses2008}%
  \BibitemOpen
  \bibfield  {author} {\bibinfo {author} {\bibfnamefont {G.}~\bibnamefont
  {Morra}}\ and\ \bibinfo {author} {\bibfnamefont {D.~A.}\ \bibnamefont
  {Yuen}},\ }\bibfield  {title} {\bibinfo {title} {Role of {{Korteweg}}
  stresses in geodynamics},\ }\href {https://doi.org/10.1029/2007GL032860}
  {\bibfield  {journal} {\bibinfo  {journal} {Geophysical Research Letters}\
  }\textbf {\bibinfo {volume} {35}},\ \bibinfo {pages} {2007GL032860} (\bibinfo
  {year} {2008})}\BibitemShut {NoStop}%
\bibitem [{\citenamefont
  {Babadagli}(2007)}]{babadagliDevelopmentMatureOil2007}%
  \BibitemOpen
  \bibfield  {author} {\bibinfo {author} {\bibfnamefont {T.}~\bibnamefont
  {Babadagli}},\ }\bibfield  {title} {\bibinfo {title} {Development of mature
  oil fields --- {{A}} review},\ }\href
  {https://doi.org/10.1016/j.petrol.2006.10.006} {\bibfield  {journal}
  {\bibinfo  {journal} {Journal of Petroleum Science and Engineering}\ }\textbf
  {\bibinfo {volume} {57}},\ \bibinfo {pages} {221} (\bibinfo {year}
  {2007})}\BibitemShut {NoStop}%
\bibitem [{\citenamefont {Swernath}\ \emph {et~al.}(2010)\citenamefont
  {Swernath}, \citenamefont {Malengier},\ and\ \citenamefont
  {Pushpavanam}}]{swernathEffectKortewegStress2010}%
  \BibitemOpen
  \bibfield  {author} {\bibinfo {author} {\bibfnamefont {S.}~\bibnamefont
  {Swernath}}, \bibinfo {author} {\bibfnamefont {B.}~\bibnamefont
  {Malengier}},\ and\ \bibinfo {author} {\bibfnamefont {S.}~\bibnamefont
  {Pushpavanam}},\ }\bibfield  {title} {\bibinfo {title} {Effect of
  {{Korteweg}} stress on viscous fingering of solute plugs in a porous
  medium},\ }\href {https://doi.org/10.1016/j.ces.2009.09.021} {\bibfield
  {journal} {\bibinfo  {journal} {Chemical Engineering Science}\ }\textbf
  {\bibinfo {volume} {65}},\ \bibinfo {pages} {2284} (\bibinfo {year}
  {2010})}\BibitemShut {NoStop}%
\bibitem [{\citenamefont {Petitjeans}\ and\ \citenamefont
  {Maxworthy}(1996)}]{petitjeansMiscibleDisplacementsCapillary1996}%
  \BibitemOpen
  \bibfield  {author} {\bibinfo {author} {\bibfnamefont {P.}~\bibnamefont
  {Petitjeans}}\ and\ \bibinfo {author} {\bibfnamefont {T.}~\bibnamefont
  {Maxworthy}},\ }\bibfield  {title} {\bibinfo {title} {Miscible displacements
  in capillary tubes. {{Part}} 1. {{Experiments}}},\ }\href
  {https://doi.org/10.1017/S0022112096008233} {\bibfield  {journal} {\bibinfo
  {journal} {J. Fluid Mech.}\ }\textbf {\bibinfo {volume} {326}},\ \bibinfo
  {pages} {37} (\bibinfo {year} {1996})}\BibitemShut {NoStop}%
\bibitem [{\citenamefont {Ban}\ \emph {et~al.}(2010)\citenamefont {Ban},
  \citenamefont {Aoyama},\ and\ \citenamefont
  {Matsumoto}}]{banSelfgeneratedMotionDroplets2010}%
  \BibitemOpen
  \bibfield  {author} {\bibinfo {author} {\bibfnamefont {T.}~\bibnamefont
  {Ban}}, \bibinfo {author} {\bibfnamefont {A.}~\bibnamefont {Aoyama}},\ and\
  \bibinfo {author} {\bibfnamefont {T.}~\bibnamefont {Matsumoto}},\ }\bibfield
  {title} {\bibinfo {title} {Self-generated {{Motion}} of {{Droplets Induced}}
  by {{Korteweg Force}}},\ }\href {https://doi.org/10.1246/cl.2010.1294}
  {\bibfield  {journal} {\bibinfo  {journal} {Chemistry Letters}\ }\textbf
  {\bibinfo {volume} {39}},\ \bibinfo {pages} {1294} (\bibinfo {year}
  {2010})}\BibitemShut {NoStop}%
\bibitem [{\citenamefont {Ban}\ \emph {et~al.}(2016)\citenamefont {Ban},
  \citenamefont {Fukuyama}, \citenamefont {Makino}, \citenamefont {Nawa},\ and\
  \citenamefont {Nagatsu}}]{banSelfPropelledVesiclesInduced2016}%
  \BibitemOpen
  \bibfield  {author} {\bibinfo {author} {\bibfnamefont {T.}~\bibnamefont
  {Ban}}, \bibinfo {author} {\bibfnamefont {T.}~\bibnamefont {Fukuyama}},
  \bibinfo {author} {\bibfnamefont {S.}~\bibnamefont {Makino}}, \bibinfo
  {author} {\bibfnamefont {E.}~\bibnamefont {Nawa}},\ and\ \bibinfo {author}
  {\bibfnamefont {Y.}~\bibnamefont {Nagatsu}},\ }\bibfield  {title} {\bibinfo
  {title} {Self-{{Propelled Vesicles Induced}} by the {{Mixing}} of {{Two
  Polymeric Aqueous Solutions}} through a {{Vesicle Membrane Far}} from
  {{Equilibrium}}},\ }\href {https://doi.org/10.1021/acs.langmuir.6b00105}
  {\bibfield  {journal} {\bibinfo  {journal} {Langmuir}\ }\textbf {\bibinfo
  {volume} {32}},\ \bibinfo {pages} {2574} (\bibinfo {year}
  {2016})}\BibitemShut {NoStop}%
%BiBlioExtraSUPMAT
\rvv{
\bibitem [{\citenamefont {Guillot}\ \emph {et~al.}(2006)\citenamefont {Guillot}, \citenamefont {Panizza}, \citenamefont {Salmon}, \citenamefont {Joanicot}, \citenamefont {Colin}, \citenamefont {Bruneau},\ and\ \citenamefont {Colin}}]{guillotViscosimeterMicrofluidicChip2006}%
  \BibitemOpen
  \bibfield  {author} {\bibinfo {author} {\bibfnamefont {P.}~\bibnamefont {Guillot}}, \bibinfo {author} {\bibfnamefont {P.}~\bibnamefont {Panizza}}, \bibinfo {author} {\bibfnamefont {J.-B.}\ \bibnamefont {Salmon}}, \bibinfo {author} {\bibfnamefont {M.}~\bibnamefont {Joanicot}}, \bibinfo {author} {\bibfnamefont {A.}~\bibnamefont {Colin}}, \bibinfo {author} {\bibfnamefont {C.-H.}\ \bibnamefont {Bruneau}},\ and\ \bibinfo {author} {\bibfnamefont {T.}~\bibnamefont {Colin}},\ }\bibfield  {title} {\bibinfo {title} {Viscosimeter on a {{Microfluidic Chip}}},\ }\href {https://doi.org/10.1021/la060131z} {\bibfield  {journal} {\bibinfo  {journal} {Langmuir}\ }\textbf {\bibinfo {volume} {22}},\ \bibinfo {pages} {6438} (\bibinfo {year} {2006})}\BibitemShut {NoStop}%
\bibitem [{\citenamefont {Cubaud}\ and\ \citenamefont {Mason}(2009)}]{cubaudHighviscosityFluidThreads2009}%
  \BibitemOpen
  \bibfield  {author} {\bibinfo {author} {\bibfnamefont {T.}~\bibnamefont {Cubaud}}\ and\ \bibinfo {author} {\bibfnamefont {T.~G.}\ \bibnamefont {Mason}},\ }\bibfield  {title} {\bibinfo {title} {High-viscosity fluid threads in weakly diffusive microfluidic systems},\ }\href {https://doi.org/10.1088/1367-2630/11/7/075029} {\bibfield  {journal} {\bibinfo  {journal} {New J. Phys.}\ }\textbf {\bibinfo {volume} {11}},\ \bibinfo {pages} {075029} (\bibinfo {year} {2009})}\BibitemShut {NoStop}%
\bibitem [{\citenamefont {De~Menech}\ \emph {et~al.}(2008)\citenamefont {De~Menech}, \citenamefont {Garstecki}, \citenamefont {Jousse},\ and\ \citenamefont {Stone}}]{demenechTransitionSqueezingDripping2008}%
  \BibitemOpen
  \bibfield  {author} {\bibinfo {author} {\bibfnamefont {M.}~\bibnamefont {De~Menech}}, \bibinfo {author} {\bibfnamefont {P.}~\bibnamefont {Garstecki}}, \bibinfo {author} {\bibfnamefont {F.}~\bibnamefont {Jousse}},\ and\ \bibinfo {author} {\bibfnamefont {H.~A.}\ \bibnamefont {Stone}},\ }\bibfield  {title} {\bibinfo {title} {Transition from squeezing to dripping in a microfluidic {{T-shaped}} junction},\ }\href {https://doi.org/10.1017/S002211200700910X} {\bibfield  {journal} {\bibinfo  {journal} {J. Fluid Mech.}\ }\textbf {\bibinfo {volume} {595}},\ \bibinfo {pages} {141} (\bibinfo {year} {2008})}\BibitemShut {NoStop}%
\bibitem [{\citenamefont {Tabeling}(2023)}]{tabelingIntroductionMicrofluidics2023}%
  \BibitemOpen
  \bibfield  {author} {\bibinfo {author} {\bibfnamefont {P.}~\bibnamefont {Tabeling}},\ }\href {https://doi.org/10.1093/oso/9780192845306.001.0001} {\emph {\bibinfo {title} {Introduction to {{Microfluidics}}}}},\ \bibinfo {edition} {2nd}\ ed.\ (\bibinfo  {publisher} {Oxford University PressOxford},\ \bibinfo {year} {2023})\BibitemShut {NoStop}%
\bibitem [{\citenamefont {Friend}\ and\ \citenamefont {Yeo}(2010)}]{friendFabricationMicrofluidicDevices2010}%
  \BibitemOpen
  \bibfield  {author} {\bibinfo {author} {\bibfnamefont {J.}~\bibnamefont {Friend}}\ and\ \bibinfo {author} {\bibfnamefont {L.}~\bibnamefont {Yeo}},\ }\bibfield  {title} {\bibinfo {title} {Fabrication of microfluidic devices using polydimethylsiloxane},\ }\href {https://doi.org/10.1063/1.3259624} {\bibfield  {journal} {\bibinfo  {journal} {Biomicrofluidics}\ }\textbf {\bibinfo {volume} {4}},\ \bibinfo {pages} {026502} (\bibinfo {year} {2010})}\BibitemShut {NoStop}%
\bibitem [{\citenamefont {Govindarajan}\ and\ \citenamefont {Sahu}(2014)}]{govindarajanInstabilitiesViscosityStratifiedFlow2014}%
  \BibitemOpen
  \bibfield  {author} {\bibinfo {author} {\bibfnamefont {R.}~\bibnamefont {Govindarajan}}\ and\ \bibinfo {author} {\bibfnamefont {K.~C.}\ \bibnamefont {Sahu}},\ }\bibfield  {title} {\bibinfo {title} {Instabilities in {{Viscosity-Stratified Flow}}},\ }\href {https://doi.org/10.1146/annurev-fluid-010313-141351} {\bibfield  {journal} {\bibinfo  {journal} {Annu. Rev. Fluid Mech.}\ }\textbf {\bibinfo {volume} {46}},\ \bibinfo {pages} {331} (\bibinfo {year} {2014})}\BibitemShut {NoStop}%
\bibitem [{\citenamefont {Govindarajan}(2004)}]{govindarajanEffectMiscibilityLinear2004}%
  \BibitemOpen
  \bibfield  {author} {\bibinfo {author} {\bibfnamefont {R.}~\bibnamefont {Govindarajan}},\ }\bibfield  {title} {\bibinfo {title} {Effect of miscibility on the linear instability of two-fluid channel flow},\ }\href {https://doi.org/10.1016/j.ijmultiphaseflow.2004.06.006} {\bibfield  {journal} {\bibinfo  {journal} {International Journal of Multiphase Flow}\ }\textbf {\bibinfo {volume} {30}},\ \bibinfo {pages} {1177} (\bibinfo {year} {2004})}\BibitemShut {NoStop}%
\bibitem [{\citenamefont {Tosun}\ \emph {et~al.}(1988)\citenamefont {Tosun}, \citenamefont {Uner},\ and\ \citenamefont {Ozgen}}]{tosunCriticalReynoldsNumber1988}%
  \BibitemOpen
  \bibfield  {author} {\bibinfo {author} {\bibfnamefont {I.}~\bibnamefont {Tosun}}, \bibinfo {author} {\bibfnamefont {D.}~\bibnamefont {Uner}},\ and\ \bibinfo {author} {\bibfnamefont {C.}~\bibnamefont {Ozgen}},\ }\bibfield  {title} {\bibinfo {title} {Critical {{Reynolds}} number for {{Newtonian}} flow in rectangular ducts},\ }\href {https://doi.org/10.1021/ie00082a034} {\bibfield  {journal} {\bibinfo  {journal} {Ind. Eng. Chem. Res.}\ }\textbf {\bibinfo {volume} {27}},\ \bibinfo {pages} {1955} (\bibinfo {year} {1988})}\BibitemShut {NoStop}%
\bibitem [{\citenamefont {Sahu}\ and\ \citenamefont {Govindarajan}(2016)}]{sahuLinearStabilityAnalysis2016}%
  \BibitemOpen
  \bibfield  {author} {\bibinfo {author} {\bibfnamefont {K.~C.}\ \bibnamefont {Sahu}}\ and\ \bibinfo {author} {\bibfnamefont {R.}~\bibnamefont {Govindarajan}},\ }\bibfield  {title} {\bibinfo {title} {Linear stability analysis and direct numerical simulation of two-layer channel flow},\ }\href {https://doi.org/10.1017/jfm.2016.346} {\bibfield  {journal} {\bibinfo  {journal} {J. Fluid Mech.}\ }\textbf {\bibinfo {volume} {798}},\ \bibinfo {pages} {889} (\bibinfo {year} {2016})}\BibitemShut {NoStop}%
\bibitem [{\citenamefont {Stiles}\ and\ \citenamefont {Fletcher}(2004)}]{stilesHydrodynamicControlInterface2004}%
  \BibitemOpen
  \bibfield  {author} {\bibinfo {author} {\bibfnamefont {P.~J.}\ \bibnamefont {Stiles}}\ and\ \bibinfo {author} {\bibfnamefont {D.~F.}\ \bibnamefont {Fletcher}},\ }\bibfield  {title} {\bibinfo {title} {Hydrodynamic control of the interface between two liquids flowing through a horizontal or vertical microchannel},\ }\href {https://doi.org/10.1039/b315524b} {\bibfield  {journal} {\bibinfo  {journal} {Lab Chip}\ }\textbf {\bibinfo {volume} {4}},\ \bibinfo {pages} {121} (\bibinfo {year} {2004})}\BibitemShut {NoStop}%
\bibitem [{\citenamefont {Stone}\ \emph {et~al.}(2004)\citenamefont {Stone}, \citenamefont {Stroock},\ and\ \citenamefont {Ajdari}}]{stoneEngineeringFlowsSmall2004}%
  \BibitemOpen
  \bibfield  {author} {\bibinfo {author} {\bibfnamefont {H.}~\bibnamefont {Stone}}, \bibinfo {author} {\bibfnamefont {A.}~\bibnamefont {Stroock}},\ and\ \bibinfo {author} {\bibfnamefont {A.}~\bibnamefont {Ajdari}},\ }\bibfield  {title} {\bibinfo {title} {Engineering {{Flows}} in {{Small Devices}}: {{Microfluidics Toward}} a {{Lab-on-a-Chip}}},\ }\href {https://doi.org/10.1146/annurev.fluid.36.050802.122124} {\bibfield  {journal} {\bibinfo  {journal} {Annu. Rev. Fluid Mech.}\ }\textbf {\bibinfo {volume} {36}},\ \bibinfo {pages} {381} (\bibinfo {year} {2004})}\BibitemShut {NoStop}%
\bibitem [{\citenamefont {D'Errico}\ \emph {et~al.}(2004)\citenamefont {D'Errico}, \citenamefont {Ortona}, \citenamefont {Capuano},\ and\ \citenamefont {Vitagliano}}]{derricoDiffusionCoefficientsBinary2004a}%
  \BibitemOpen
  \bibfield  {author} {\bibinfo {author} {\bibfnamefont {G.}~\bibnamefont {D'Errico}}, \bibinfo {author} {\bibfnamefont {O.}~\bibnamefont {Ortona}}, \bibinfo {author} {\bibfnamefont {F.}~\bibnamefont {Capuano}},\ and\ \bibinfo {author} {\bibfnamefont {V.}~\bibnamefont {Vitagliano}},\ }\bibfield  {title} {\bibinfo {title} {Diffusion {{Coefficients}} for the {{Binary System Glycerol}} + {{Water}} at 25 {$^\circ$}{{C}}. {{A Velocity Correlation Study}}},\ }\href {https://doi.org/10.1021/je049917u} {\bibfield  {journal} {\bibinfo  {journal} {Journal of Chemical \& Engineering Data}\ }\textbf {\bibinfo {volume} {49}},\ \bibinfo {pages} {1665} (\bibinfo {year} {2004})}\BibitemShut {NoStop}%
\bibitem [{\citenamefont {Kjelstrup}\ and\ \citenamefont {Bedeaux}(2008)}]{kjelstrupNonequilibriumThermodynamicsHeterogeneous2008}%
  \BibitemOpen
  \bibfield  {author} {\bibinfo {author} {\bibfnamefont {S.}~\bibnamefont {Kjelstrup}}\ and\ \bibinfo {author} {\bibfnamefont {D.}~\bibnamefont {Bedeaux}},\ }\href@noop {} {\emph {\bibinfo {title} {Non-Equilibrium Thermodynamics of Heterogeneous Systems}}},\ \bibinfo {series} {Series on Advances in Statistical Mechanics}\ No.\ \bibinfo {number} {Volume 16}\ (\bibinfo  {publisher} {World Scientific},\ \bibinfo {address} {Singapore},\ \bibinfo {year} {2008})\BibitemShut {NoStop}%
\bibitem [{\citenamefont {Orza}\ \emph {et~al.}(1997)\citenamefont {Orza}, \citenamefont {Brito}, \citenamefont {Van~Noije},\ and\ \citenamefont {Ernst}}]{orzaPatternsLongRange1997}%
  \BibitemOpen
  \bibfield  {author} {\bibinfo {author} {\bibfnamefont {J.~A.~G.}\ \bibnamefont {Orza}}, \bibinfo {author} {\bibfnamefont {R.}~\bibnamefont {Brito}}, \bibinfo {author} {\bibfnamefont {T.~P.~C.}\ \bibnamefont {Van~Noije}},\ and\ \bibinfo {author} {\bibfnamefont {M.~H.}\ \bibnamefont {Ernst}},\ }\bibfield  {title} {\bibinfo {title} {Patterns and {{Long Range Correlations}} in {{Idealized Granular Flows}}},\ }\href {https://doi.org/10.1142/S0129183197000825} {\bibfield  {journal} {\bibinfo  {journal} {Int. J. Mod. Phys. C}\ }\textbf {\bibinfo {volume} {08}},\ \bibinfo {pages} {953} (\bibinfo {year} {1997})},\ \Eprint {https://arxiv.org/abs/cond-mat/9702029} {arXiv:cond-mat/9702029} \BibitemShut {NoStop}%
\bibitem [{\citenamefont {Guazzelli}\ and\ \citenamefont {Hinch}(2011)}]{guazzelliFluctuationsInstabilitySedimentation2011}%
  \BibitemOpen
  \bibfield  {author} {\bibinfo {author} {\bibfnamefont {{\'E}.}~\bibnamefont {Guazzelli}}\ and\ \bibinfo {author} {\bibfnamefont {J.}~\bibnamefont {Hinch}},\ }\bibfield  {title} {\bibinfo {title} {Fluctuations and {{Instability}} in {{Sedimentation}}},\ }\href {https://doi.org/10.1146/annurev-fluid-122109-160736} {\bibfield  {journal} {\bibinfo  {journal} {Annu. Rev. Fluid Mech.}\ }\textbf {\bibinfo {volume} {43}},\ \bibinfo {pages} {97} (\bibinfo {year} {2011})}\BibitemShut {NoStop}%
\bibitem [{\citenamefont {Katsuragi}\ \emph {et~al.}(2010)\citenamefont {Katsuragi}, \citenamefont {Abate},\ and\ \citenamefont {Durian}}]{katsuragiJammingGrowthDynamical2010}%
  \BibitemOpen
  \bibfield  {author} {\bibinfo {author} {\bibfnamefont {H.}~\bibnamefont {Katsuragi}}, \bibinfo {author} {\bibfnamefont {A.~R.}\ \bibnamefont {Abate}},\ and\ \bibinfo {author} {\bibfnamefont {D.~J.}\ \bibnamefont {Durian}},\ }\bibfield  {title} {\bibinfo {title} {Jamming and growth of dynamical heterogeneities versus depth for granular heap flow},\ }\href {https://doi.org/10.1039/b918991b} {\bibfield  {journal} {\bibinfo  {journal} {Soft Matter}\ }\textbf {\bibinfo {volume} {6}},\ \bibinfo {pages} {3023} (\bibinfo {year} {2010})}\BibitemShut {NoStop}%
\bibitem [{\citenamefont {Wu}\ \emph {et~al.}(1991)\citenamefont {Wu}, \citenamefont {Pine},\ and\ \citenamefont {Dixon}}]{wuEnhancedConcentrationFluctuations1991}%
  \BibitemOpen
  \bibfield  {author} {\bibinfo {author} {\bibfnamefont {X.-L.}\ \bibnamefont {Wu}}, \bibinfo {author} {\bibfnamefont {D.~J.}\ \bibnamefont {Pine}},\ and\ \bibinfo {author} {\bibfnamefont {P.~K.}\ \bibnamefont {Dixon}},\ }\bibfield  {title} {\bibinfo {title} {Enhanced concentration fluctuations in polymer solutions under shear flow},\ }\href {https://doi.org/10.1103/PhysRevLett.66.2408} {\bibfield  {journal} {\bibinfo  {journal} {Phys. Rev. Lett.}\ }\textbf {\bibinfo {volume} {66}},\ \bibinfo {pages} {2408} (\bibinfo {year} {1991})}\BibitemShut {NoStop}%
  } %This parenthesis closes \rvv
\end{thebibliography}

%

\end{document}